\newcommand{\DeltaPDC}{\Delta^{\rm PDC}}                   
\newcommand{\DeltaSFG}{\Delta }                   
\newcommand{\cV}{{\cal V}_{\rm PDC}}
\newcommand{\cW}{\Phi_{\rm SFG}}

\newcommand{\vka}{\vec{w}}			
\newcommand{\w}{\vec{w}}			
	
\newcommand{\vkap}{\vec{w}\,'}

\newcommand{\sinc}{{\rm sinc}}
\newcommand{\q}{\vec{q}}

\newcommand{\nn}{\nonumber}
\newcommand{\bsub}{\begin{subequations}}
\newcommand{\esub}{\end{subequations}}
\newcommand{\beq}{\begin{equation}}
\newcommand{\eeq}{\end{equation}}
\newcommand{\beqa}{\begin{eqnarray}}
\newcommand{\eeqa}{\end{eqnarray}}
\newcommand{\beql}{\begin{subequations}\begin{eqnarray}}
\newcommand{\eeql}{\end{eqnarray}\end{subequations}}
\documentclass[osajnl,twocolumn,showpacs,superscriptaddress,nofootinbib,10pt]{revtex4-1} 
\usepackage{amsmath,amssymb,graphicx}
\usepackage{float}   
\usepackage{verbatim}  
\usepackage{braket}
\begin{document}
\title{Space-time coupling in the up-conversion of broadband down-converted light}
\author{Enrico Brambilla}  
\affiliation{Dipartimento di Scienza e Alta Tecnologia\text{,} Universit\`a dell'Insubria, Via Valleggio 11 22100 Como, Italy}
\author{Ottavia Jedrkiewicz} 
\affiliation{CNR, Istituto di Fotonica e Nanotecnologie, Piazza Leonardo da Vinci 32, 20133 Milano, Italy}
\author{Paolo Di Trapani} 
\affiliation{Dipartimento di Scienza e Alta Tecnologia\text{,} Universit\`a dell'Insubria, Via Valleggio 11 22100 Como, Italy}
\author{Alessandra Gatti}
\affiliation{CNR, Istituto di Fotonica e Nanotecnologie, Piazza Leonardo da Vinci 32, 20133 Milano, Italy}
\affiliation{Corresponding author: Alessandra.Gatti@mi.infn.it}
\begin{abstract}
We investigate the up-conversion process of broadband light from parametric down-conversion (PDC), focusing on the spatio-temporal spectral properties of the 
sum-frequency generated (SFG) radiation.
We demonstrate that the incoherent component of the SFG spectrum is characterized by a skewed  geometry in space-time, which originates  from a compensation between the group-velocity mismatch and the spatial walk-off of the fundamental and the SFG fields.
The results are illustrated both by a theoretical modeling of the optical system and by experimental measurements.
\end{abstract}
\maketitle
\section*{Introduction}
\label{sec:intro}
The process of sum frequency generation (SFG) occurring in a $\chi^{(2)}$ crystal has often been used for probing the spatio-temporal structure of ultrafast femtosecond pulses \cite{trebinobook}. In particular, the three-dimensional mapping of ultrashort complex
pulses \cite{minardi2004,trull2004} and the observation of the spatio-temporal dynamics of Kerr media filamentation \cite{majus2010} has been obtained with this method. Recent works \cite{dayan2005,peer2005b,ribeiro2006,harris2007,dayan2007} have shown
that sum-frequency generation  also represents a resource for exploring the temporal entanglement  of twin photons or twin beams generated by parametric down-conversion (PDC), and for applications in quantum information processing.
For example, spectrally engineered SFG has been proposed as a tool for selecting quantum Schmidt modes 
from the broadband  state generated by a periodically poled Lithium Niobate waveguide \cite{silberhorn2011}.
\par
Recently,  our  group used the SFG process for investigating the quantum correlation of PDC twin beams in the whole spatio-temporal domain 
\cite{brambilla2012,jedr2011, jedr2012a,jedr2012b}, and demonstrated its X-shaped geometry, non-factorable in space-time  \cite{gatti2009,caspani2010,brambilla2010, gatti2012}. In that experiment,  twin beams generated from a first $\chi^{(2)}$ crystal were injected in a second identical crystal where up-conversion took place. We showed that the inverse process of PDC, i.e. the {\em coherent} up-conversion of twin photons belonging to phase conjugated modes, allows to reconstruct the spatio-temporal correlation of twin beams, as  a function of a controlled temporal delay and spatial transverse diplacement between the twin beams \cite{brambilla2012, jedr2012b}. 
\par
In this work we shall instead focus on the spatio-temporal properties of the {\em incoherent}  component of the up-converted  light, which originates from the
random up-conversion of photon pairs  not belonging to twin modes.
These incoherent processes, though negligible compared to the coherent ones in the coincidence count regime, become relevant 
at high parametric gains, and give rise to a very broadband incoherent component. 
We shall investigate the spatio-temporal properties of this incoherent component, and show
that propagation along the SFG crystal rapidly selects the up-converted spatio-temporal modes, 
leading to a coupling between the spatial and the temporal frequencies. Correspondingly, 
the SFG spectrum displays a skewed geometry in the spatio-temporal domain, which we shall interpret  in terms of an interplay of the  group-velocity mismatch and transverse spatial walk-off between the fundamental and the up-converted fields. 
\par
These features of the SFG spectrum are similar in nature to the effects observed in the up-conversion of ultrafast optical pulses, when thick crystals are used \cite{oshea2001, wasilewsky2004, branderhorst2006}, for example in the measurement of ultrafast pulses by means of the GRENOUILLE technique \cite{oshea2001,trebinobook}. In this technique the various wavelengths of the upconverted light undergo an angular separation when propagating in a thick crystal, when the input pulse is focused in one transverse direction. 
In our case, similar effects can be observed (perhaps over a much wider range of temporal and spatial frequencies) because the input PDC light is naturally extremely broadband both in the temporal and spatial domain.  
\par
Notice that similar mechanisms,  where the space-time coupling plays a role, have been described for various nonlinear
processes. We can cite the interplay of spatial diffraction and temporal dispersion that is at the
origin of the so-called X-waves \cite{couairon2006,ditrapani2003} and of the X-shaped coherence \cite{jedr2006} and correlation of twin beams \cite{gatti2009,caspani2010,brambilla2010,gatti2012}. 
\par
In the first part of this work (Sec.\ref{sec:model}) we discuss theoretically the process of incoherent upconversion of PDC light: our model considers the temporal and spatial degrees of freedom of light on the same ground, which allows us to develop a clear description  of the physical mechanism at the origin of the skewed geometry of the SFG spectrum. A short comparison with the upconversion of an ultrafast pulse will outline similarities and differences between the two cases.
The second part of the paper (Sec.\ref{sec:exp}) will present the results of an experimental observation of these features,  and a comparison with the theory. 
%
%
%
\section{Optical setup}
\label{sec:exp_scheme}
We start with a description of the experimental setup, whose results will be presented  in  section \ref{sec:exp}.
\par
The setup is similar to that reported in \cite{jedr2011} and is illustrated in Fig.1.  
\begin{figure}[b,h]
\centering
{\scalebox{0.45}{\includegraphics*{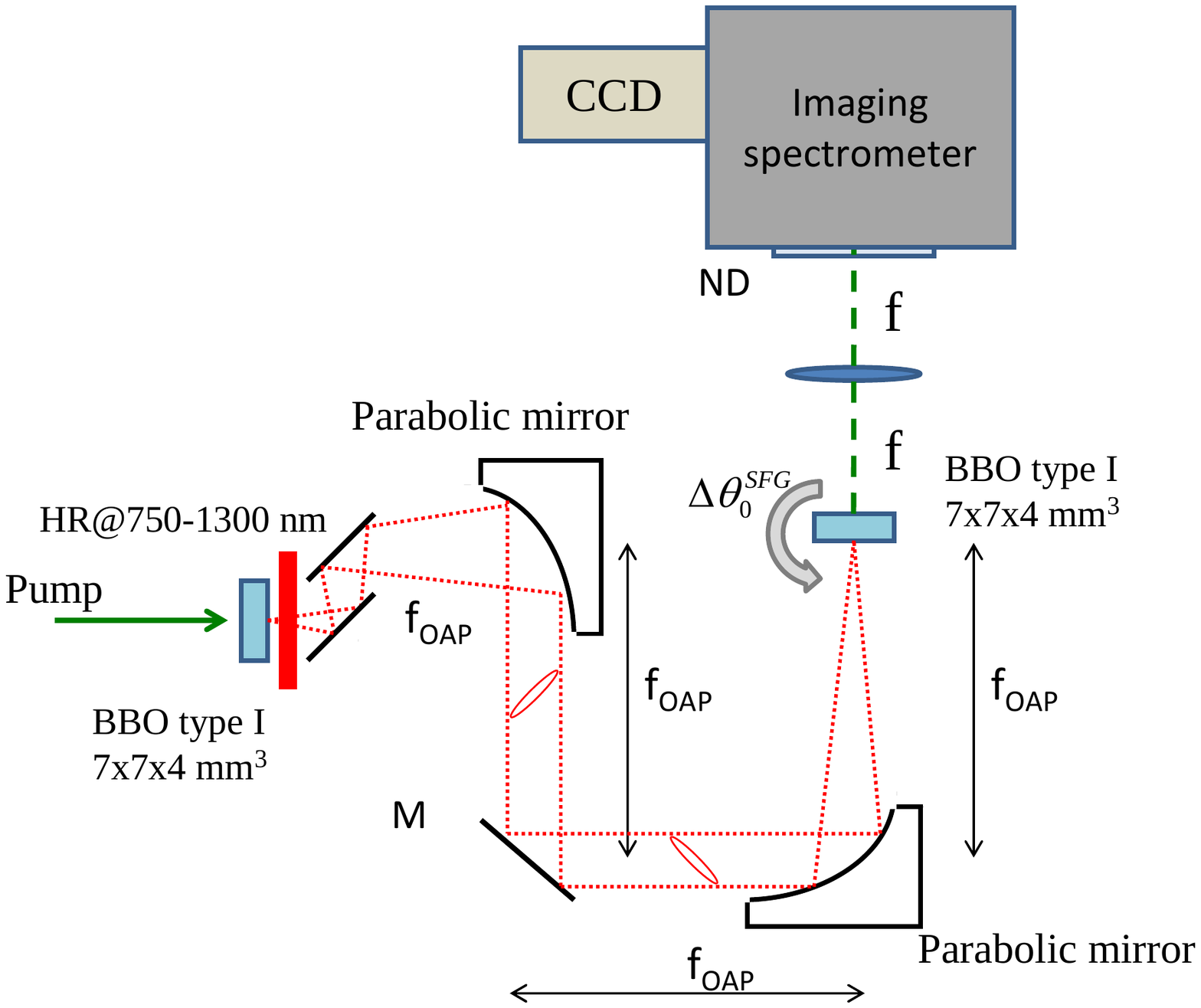}}}
\caption{(color online) Experimental setup. The output face of a PDC crystal is imaged onto the input face of  an identical SFG crystal, which is mounted on a micrometric rotation stage that allows a fine tuning of the phase-matching conditions.
The SFG far-field, observed in the focal plane of a lens, is analyzed by an imaging spectrometer.}
\label{fig_setup}
\end{figure}
The PDC radiation is generated by a type I BBO crystal pumped at $\lambda_0=527.5$nm,  in a pulsed regime with high parametric gain.   Precisely, the pump beam  is a $\sim 1$ps  pulse with a $\sim 700$ micrometers transverse beam waist.
The crystal is cut for collinear emission at the degenerate wavelength $2\lambda_0=1055$nm. After the PDC crystal,
the pump is  subtracted by a glass filter (transmission bandwidth 750-1300nm) 
and the PDC field is then imaged at the entrance face of a second  identical crystal where the up-conversion process takes place.
The 4-f imaging device is built with two achromatic parabolic mirrors rather than with lenses, in order to minimize dispersion 
between the two crystals. As demonstrated in \cite{jedr2011,jedr2012a}, it is capable of a careful imaging of  the PDC field
at the SFG crystal input face over a very broad range of temporal and spatial frequencies,  on the order of $\Delta \lambda\sim 600$ nm and $\Delta \alpha \sim \pm 4^{\circ}$, respectively . 
\par
The spatio-temporal features of the output field from the SFG crystal are analyzed in its far-field, by means of
an imaging spectrometer that collects the photons 
in the focal plane of a 20 cm focal length lens. A scientific camera (charged coupled device, Roger Scientific) 
is placed at the very output of the spectrometer (LOT Oriel) in order to record the resulting spectral images.
(see Fig.\ref{fig_setup}).
\par
For our type I BBO crystals, both the pump injected in the first crystal and the SFG light in the second crystal  are extraordinarily polarized while the PDC field is ordinarily polarized. 
As shown in Fig.2, we take into account the possibility that the two crystals are not perfectly tuned, i.e. we consider the possibility that the orientation angles of the two crystal axes with respect to the mean direction of propagation of the pump (z-axis in Fig.2),
denoted by $\theta_0^{\rm PDC}$ and $\theta_0^{\rm SFG}$, are slightly different.  
\begin{figure}
\centering
{\scalebox{0.52}{\includegraphics*{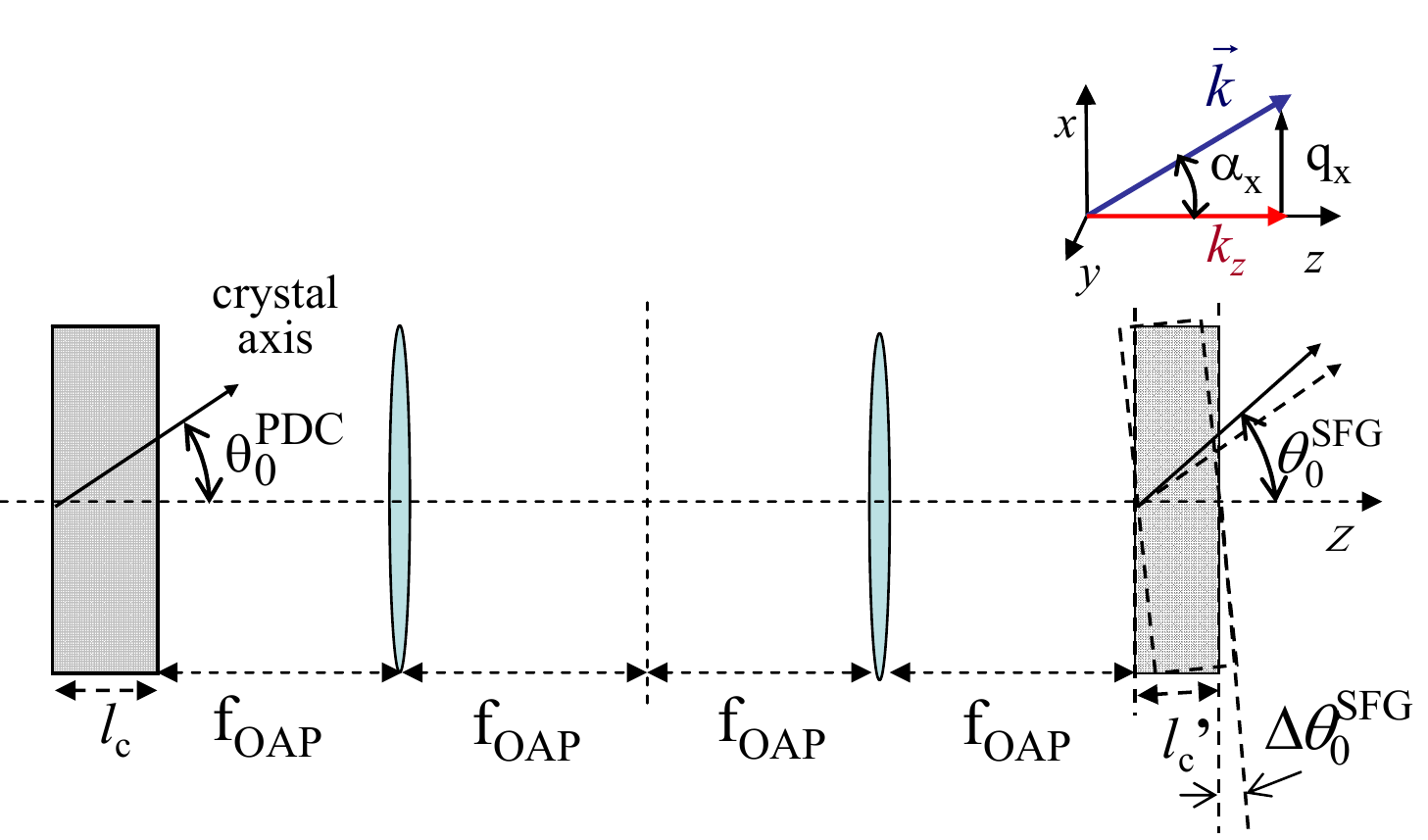}}}
\caption{(color online) Unfolded and simplified view of the setup of Fig.\ref{fig_setup}. The 4-f telescopic system images the PDC source onto the second
crystal. $\theta_0^{PDC}$ and $\theta_0^{SFG}$ are the angles formed by the two crystal axes with the pump propagation direction. 
The tilt angle $\Delta\theta_0^{SFG}=\theta_0^{SFG}-\theta_0^{PDC}$ is largerly exaggerated.}
\label{fig_scheme}
\end{figure}
\section{Theoretical modeling}
\label{sec:model}
This section provides a theoretical description of the above scheme. Our model for the up-conversion of twin beams has been developed  in  detail in Ref.\cite{brambilla2012}. In the following  we shall report only the results useful to our target, which is  the description of the incoherent upconversion spectrum. 
\par
The model is fully 3D+1, i.e. it includes  time and the two spatial coordinates in the plane transverse to the mean propagation direction of light $z$ (see Fig.2). For briefness of notation we use the 3D vector $\vka\equiv(\q,\Omega)$ in order to denote a Fourier mode of the light of temporal frequency $\omega_j+\Omega$ and transverse component of the wave-vector $\q=(q_x,q_y)$, where the reference frequencies $\omega_j$ is either the high frequency $\omega_0$,  when dealing with pump or the  up-converted field $a_0$,
or the frequency $\omega_1=\omega_0/2$ when dealing with the down-converted field $a_1$.
\par
The model is written in terms of propagation equations along the two crystals for the field operators associated with the two wavepackets of central frequencies $\omega_1$  and $\omega_0$: 
\bsub
\beqa
\frac{d}{dz} a_1(\w,z) &=& \sigma \int \frac {d^3 \w_0 }{(2\pi)^{3/2}} \left[ 
a_0 (\w_0,z) a_1^{\dagger} (\w_0-\w,z) \right. \nonumber \\
& & \left. e^{-i \Delta (\w, \w_0-\w) z} \right] \, ,
\label{prop1} \\
\frac{d}{dz} a_0(\w_0,z) &=& -\sigma \int \frac {d^3 \w }{(2\pi)^{3/2}} 
\left[a_1 (\w,z) a_1 (\w_0-\w,z) \right. \nonumber \\
& & \left. e^{i \Delta (\w, \w_0-\w) z}\right]
\label{prop2} 
\eeqa 
\esub
where the coupling constant $\sigma$ is proportional to the $\chi^{(2)}$ coefficient of the nonlinear medium, and we introduced the phase matching function 
\beq
\label{delta}
\Delta(\vka,\vka_0 -\w)=k_{1z}(\w)+k_{1z}(\w_0 -\w)-k_{0z}(\w_0)\;,
\eeq
where $k_{jz}(\vka)=\sqrt{k_j(\vka)^2-q^2}$, $j=0,1$ are the longitudinal components of the wave vectors $\vec{k}_j(\w)$. 
When considering the PDC process, the field $a_1$ at the entrance face of the first crystal is taken in the vacuum state, while the pump field $a_0$ is described as an intense coherent field. When considering the SFG process, the field  $a_0$ at the entrance face of the second crystal  is taken  in the vacuum state, while the fundamental field $a_1$ is exactly the output from the first crystal (the 4f system is treated as a perfect imaging system).   
\par
Our description has two levels of investigation. At the first level, analytical or semy-analytical results can be derived by  approximating  the pump beam driving the PDC process by a plane-wave monochromatic beam. This may seem a rather rough approximation, but in fact turns out quite valid in our experimental conditions. At the second level, numerical simulations as those described in Refs. \cite{brambilla2004,brambilla2012}) can account for the finite size and duration of the pump pulse. The numerical simulations perform a full 3D+1 modeling of the propagation equations along the two crystal, where the generation of the PDC field is described by a stochastic method in the framework of the
Wigner representation and 
the phase-matching conditions in both crystals are described by using the complete Sellmeier dispersion relation \cite{boeuf2000} for a BBO crystal.
\subsection {Plane-wave pump model} 
In the remaining of this section we shall consider the limit where the pump beam driving the PDC can be approximated by a monochromatic plane-wave  propagating 
along the $z$-axis.
As analyzed in detail in \cite{caspani2010}, such an approximation holds as long as the pump beam waist and duration are 
larger than the spatial transverse displacement and temporal delay, respectively,  experienced by the pump and signal beams along the crystal because of spatial walk-off
and group velocity dispersion.  For the 4mm long BBO crystal employed in the experiment, a pump pulse with a waist larger than $\sim 250\,\mu$m and a duration above $\sim 350\,$fs satisfies these conditions.
\par
In the first crystal, a roughly monochromatic and plane-wave pump populates only the mode $\vka =0$, so that the only processes allowed are those where a  pump photon in mode $\vka=0$ is down-converted 
into pairs of photons belonging to phase-conjugated modes, $\vka$
and $-\vka$, because of transverse momentum and energy conservation.  The efficiency of each of these processes is ruled by the longitudinal phase matching \eqref{delta}, which in this limit reduces to: 
\beq
\label{delta_PW}
\DeltaPDC(\vka,-\vka)=k_{1z}(\vka)+k_{1z}(-\vka)-k_0,
\eeq
where $k_0 \equiv k_{0z}(\vka=0)$is the wave-number of the pump.  
\par
In the same limit,  the PDC spectral intensity can be expressed as  \cite{gatti2003,brambilla2004}
\label{PDC_spectrum}
\beqa
&&{\cal I}_{\rm PDC}(\vka)= \frac{g^2\sinh^2 \sqrt{g^2-\frac{1}{4}\left[\DeltaPDC(\vka,-\vka)l_c\right]^2} }
                                                  {g^2-\frac{1}{4}\left[\DeltaPDC(\vka,-\vka)l_c\right]^2}
\label{V}
\eeqa
and takes its maximum value in the region of $\vka$-space where the PDC phase-mismatch $\DeltaPDC(\vka,-\vka)=0$ (see Fig.\ref{fig_inc}a
for a plot of ${\cal I}_{\rm PDC}(\vka)$).
The dimensionless parametric gain $g$, proportional to the pump amplitude, the PDC crystal length $l_c$ and the coupling coefficient $\sigma$, determines the number of photons $\sinh^2 g$ for the perfectly phase-matched mode pairs.
\par
Two different kinds of up-conversion processes take place inside the second crystal. 
The first one corresponds to the inverse PDC process and involves pairs of photons in phase-conjugated
modes, $\vka$ and $-\vka$, which are coherently up-converted into the original $\vka=0$ pump mode. 
These coherent processes lead to the partial reconstruction of the original pump beam and have been investigated in \cite{jedr2011,dayan2007}.
The second kind of processes gives rise to the incoherent SFG component, and
consists in the up-conversion of photons not belonging to phase-conjugated mode pairs, say $\vka$ and $\vkap$,
into a SFG mode $\vka+\vkap$ which differs from the pump mode. These up-conversion processes become 
relevant in the stimulated PDC regime we are considering (i.e. for $g>1$),
and generate a broadband incoherent field which, when observed in the far field of the SFG source, appears as 
a speckle-like background that spreads around the strongly focused coherent component \cite{jedr2011}.  
Its spatio-temporal spectrum is found to be (see \cite{brambilla2012} for a derivation)
\beqa
{\cal I}_{\rm SFG}(\vka_0)
&=& 2(\sigma l_c')^2\int\frac{d\vka}{(2\pi)^{3}} \left[ {\cal I}_{\rm PDC}(\vka) 
 {\cal I}_{\rm PDC}(\vka_0-\vka) \right. \nn \\
& \times& \left. 
 \left| F_{\rm SFG} (\vka, \vka_0-\vka)    \right|^2 \right] 
\label{S_PWPA}
\eeqa
where $l_c'$ is the length of the SFG crystal, ${\cal I}_{\rm PDC}(\vka)$ is the spectral intensity of the  PDC light injected into the SFG crystal ( Eq.(\ref{PDC_spectrum}), and 
\beqa
F_{\rm SFG} (\vka, \vka_0-\vka) = \sinc { \frac{\DeltaSFG (\vka, \vka_0-\vka)l_c'}{2} } \, e^{ i \frac{ \DeltaSFG(\vka, \vka_0-\vka)l_c'}{2} }   
\label{FSFG}
\eeqa
 represents the probability amplitude of  up-converting photons from the modes    $\vka, \vka_0-\vka$ into the mode  $\vka_0$, and is basically a filter based on the phase matching function in the second crystal $\DeltaSFG $ (given by the general espression \eqref{delta}). 
According to Eq.\eqref{S_PWPA}, the SFG intensity in a given mode $\vka_0$ is the incoherent sum  of all the contributions 
from the mode pairs $\vka$ and $\vka_0-\vka$ allowed by the transverse momentum and energy conservation: it is the sum 
of the product of their intensities 
weighted by the corresponding up-conversion probability   $\propto\sinc^2\frac{1}{2}\left[\Delta(\vka, \vka_0-\vka)l_c'\right]$.
\par
In the limiting case where the SFG crystal is extremely thin, the SFG spectrum reduces to the self-convolution of the PDC spectrum, since the 
SFG probability amplitude  can be replaced by unity under the integral sign in Eq.(\ref{S_PWPA}) when $l_c'\ll l_c$. 
We shall verify, however, that propagation effects inside the SFG crystal become relevant as soon as realistic propagation lengths in the SFG crystal are considered, giving the SFG incoherent spectrum a peculiar skewed geometry in the spatio-temporal frequency domain which highlights
the interplay between the spatial and temporal degree of freedoms.
\par
Although expression (\ref{S_PWPA}) has been derived for the case of a PDC source  within a quantum formalism \cite{brambilla2012}, 
it is worth noticing that it holds for any classical source of incoherent light with a field coherence function of the form 
$\langle a^*(\vka)a(\vkap)\rangle=\delta(\vka-\vkap){\cal I}_S(\vka)$, ${\cal I}_S(\vka)$ denoting the spectral density of the source.
\subsection {Analogy with the up-conversion of an ultrafast optical pulse} 
Up-conversion of ultrafast pulses in thick crystals has been analysed by several Authors (see e.g \cite{oshea2001, wasilewsky2004, branderhorst2006}. 
For the sake of simplicity, we consider here a completely coherent pulse, that is, a transform limited pulse,  of central frequency $\omega_0/2$,  impinging on a SFG crystal 
\footnote{Notice that the in the most common configuration two replicas of the same pulse impinge on the crystal at some crossing angle. We shall  keep the analogy with the PDC light by considering upconversion of a single pulse containing several angles of propagation}. 
We assume a well focused ultrafast pulse, with a  wide range of temporal frequencies and propagation directions. 
In the framework of our spatio-temporal formalism, we denote by $\alpha_1(\w)$ the classical field amplitude of the input pulse in the spatio temporal Fourier domain $\w=(\q,\Omega)$.  
The main formula that gives  
the amplitude of the up-converted field of central frequency $\omega_0$ in mode $(\w_0)$ reads:  
\beq
\alpha_0 (\w_0) = \int \frac{d \vka}{(2\pi)^{3 \over 2}}   \; \alpha_1 (\w) \alpha_1 (\w_0-\w) F_{\rm SFG} (\w,\w_0 -\w) \, .
\label{ultrafast}
\eeq
where $F_{\rm SFG}$ is the same upconversion probability amplitude given by Eq. \eqref{FSFG}. 
\par
Analogies and differences with the incoherent upconversion of PDC light can be straightfowardly observed. Equation \eqref{ultrafast} describes coherent up-conversion processes, and can be read as a coherent superposition of the {\em probability amplitudes} of all the processes by which photon pairs in the spatio-temporal modes $\w$ and $\w_0-\w$ of the input pulse are upconverted in the output SFG mode $w_0$. 
Eq. \eqref{S_PWPA} describes instead an {\em incoherent} sum of the probabilities of all the possible upconversion events. 
In both cases, the sum over the possible processes is strongly filtered by the phase matching in the SFG crystal 
In the limit of a very thin crystal, Eq.\eqref{ultrafast} reduces to the self-convolution of the pulse spatio-temporal spectral amplitude (its temporal version is typically the base of the FROG technique, see e.g \cite{trebinobook}). 
In the same limit the incoherent espression \eqref{S_PWPA} reduces to the self-convolution of the input spatio-temporal spectrum. 
However, for a thick crystal, in both cases effects arising from phase matching in the SFG crystal have to be considered: when the input light has a broad enough spectrum in both the temporal a spatial domain  these effects, as we shall sees in the next section, involve a coupling between the  spatial and temporal degrees of freedom.  
\subsection{The phase matching mechanism in the up-conversion of broadband light}
This rather long section introduces some approximations of the phase matching function \eqref{delta},  which will be useful to enlight the physical mechanism of upconversion of broadband light. 
\par
We start by making a Taylor espansion of the longitudinal wave-vectors $k_{jz}(\vka)=\sqrt{k_j^2(\q,\Omega)-q^2}$ entering into Eq. \eqref{delta}: 
\bsub
\label{k01z_quad}
\beqa
&&k_{1z}(\q,\Omega)
\approx k_1+k_1'\Omega+\frac{1}{2}k_1''\Omega^2-\frac{q^2}{2k_1}...\;,\label{k1z_quad}\\
&&k_{0z}(\q,\Omega)
\approx k_0 +k_0\,'\Omega+\frac{1}{2}k_0''\Omega^2 + \frac{\partial k_0}{\partial q_x} \, q_x-\frac{q^2}{2k_0}....,
\label{k0z_quad}
\eeqa
\esub
where $k_j\equiv k_j(\q=0, \Omega=0)$,
$k_j' = \partial k_j/ \partial\Omega |_{\q=0, \Omega=0}$,  $k_j'' = \partial^{2} k_1/ \partial\Omega^{2} |_{\q=0,\Omega=0}$, etc. 
Thus in Eqs. \eqref{k01z_quad} the zero order terms are the wave mumbers of the two waves calculated at the reference frequencies, when they propagate collinearly along the z-direction: $k_j = n_j (\omega_j, \q=0) \omega_j/c$. The terms linear in  $\Omega$ account for   the group velocities of the two waves at the central frequencies
$v_{g j} = 1/{k'_j}$ , the terms quadratic in $\Omega$ describe the effect of the dispersion of the group velocities (GVD), the terms quadratic in $q$ account for the spatial diffraction in the paraxial approximation. 
A term linear in $q_x$ is present only in Eq.\eqref{k0z_quad} for the upconverted wave. At first order in $q/k_0$, it accounts 
for the dependence of the index of refraction of the extraordinary wave on its propagation direction, that is,  on the angle between its wave-vector $\vec{k}_0$ and the crystal axis. This in turn can be espressed in terms of the angle $\alpha_x$ formed by $\vec{k}_0$ and the $z$-axis, in the plane $(x,z)$ containing the crystal axis (see Fig.2).   Physically,  it describes the effect of the spatial walk-off of the Poynting vector of the extraordinary wave,  
\beq 
\label{rho0}
\rho_0 := -\frac{1}{n_0} \frac {\partial n_0 } {\partial \alpha_x} \left. \right|_{q=0, \Omega=0}
\approx - \frac{\partial k_0}{\partial q_x} 
\eeq
being the {\em walk-off angle}.
\par
The espansions \eqref{k01z_quad}are then inserted into Eq.\eqref{delta} for the phase mismatch. Let us now assume 
that the spatio-temporal bandwidth of the input light is small enough, so that the quadratic terms can be neglected with respect to to the linear ones ({\em linear approximation} of the phase mismatch). In thhis way, we obtain an espression that depends only upon the frequency and transverse wave-vector of the up-converted field:
\beq
\DeltaSFG(\vka,\vka_0- \vka) 
\approx (2k_1- k_0) 
 + \left( \frac{\Omega_0 }{v_{g1}} -\frac{\Omega_0 }{v_{g0}}\right) 
+ \rho_0 q_{0x}
\eeq
i.e.
\beq
\DeltaSFG(\vka,\vka_0- \vka) l_c^\prime
= (2k_1- k_0) l_c^\prime -\frac {\Omega_0} {\Omega_{\rm GVM}} + \frac {q_{0x}}{q_{\rm WO} }\, ,
\label{delta_linear}
\eeq
where  we introduced the two bandwidths associated with the group velocity mismatch (GVM) and the spatial transverse walk-off: 
\beqa
\Omega_{\rm GVM}
&=& \frac{1}{ \tau_{\rm GVM} }\, ,
\quad 
\tau_{\rm GVM} = \frac{l_c'} {v_{g0}}   -\frac{l_c'} {v_{g1}}        
\label{omg_GVM} \\
q_{\rm WO}
&=& \frac{1} {l_{\rm WO} }\, ,
\quad l_{\rm WO} = {\rho_0 l_c'} \, .
\label{q_SW}
\eeqa
Here $\tau_{\rm GVM}$ and $l_{\rm WO}$  are the temporal delay and the transverse separation between the second harmonic and the fundamental fields, acquired because of the temporal and spatial walk-off during propagation. 
Espression \eqref{delta_linear} shows an angular dispersion of the upconverted frequencies: in order to satisfy phase matching, different temporal frequencies propagate with different trasverse wave-vectors (i.e. at different angles), provided that the input light has an angular spectrum sufficiently wide, so to provide enough range of up-converted wave-vectors $q_{0x}$. 
This effect is well-known in the framework of up-conversion of ultrafast pulses in thick crystals, and is esploited in the GRENOUILLE technique \cite{oshea2001, trebinobook} in order to separate angularly the various frequencies of the SFG light. Typically, in the literature (see e.g \cite{wasilewsky2004, oshea2001}, espressions similar to \eqref{delta_linear} are written in the purely temporal domain. Here, we  write a result that includes space and time on the same ground, and we  interpret the angular separation of temporal frequencies as an effect do a balance between the group velocity mismatch and the spatial walk-off during propagation along a thick SFG crystal. 
\par
The linear approximation \eqref{delta_linear}  of phase matching is quite general, and holds as long as the fundamental beam
has a small enough bandwidth. For the temporal terms, it requires that the effects of GVD are negligible over the entire temporal bandwidth:
\beqa
\frac{\Omega_{max}} {\Omega_{\rm GVD}} &\ll& \pi  \quad {\Omega_{\rm GVD}} := \frac{1}{\sqrt{k''_1 l'_c}} \, ,
\label{omegamax} 
\eeqa
where $\Omega_{max}$ is  the largest frequencies present in the spectra of the input light, and we introduced the group velocity dispersion bandwidth. 
\begin{table}[h]
  \begin{center}
    \begin{tabular}{lcl p{1.7 cm}}
    \hline \vspace{0.2 cm}
    $\Omega_{\rm GVM}$ \; &  $ \left(\frac{l_c}{v_{g0}}- \frac{l_c}{v_{g1}}\right)^{-1}  \; $  
                                            & $3 \times 10^{12} s^{-1}$ \;  & GVM bandwidth \\
    \hline
    $\Omega_{\rm GVD}$   & $\frac{1}{ \sqrt{k_1'' l_c}  }  $  & $7.6 \times 10^{13} s^{-1}$ \;    
						& Dispersion bandwidth@$\frac{\omega_0}{2}$ \\
\hline
    $\Omega_{0,\rm GVD}$   & $\frac{1}{ \sqrt{k_o'' l_c}  } $  & $ 4.4 \times 10^{13} s^{-1}$ \;    
								 & Dispersion bandwidth@$\omega_0$  \\
\hline
    $q_{\rm WO}$ & $\frac{1}{\rho_0 l_c}$ & $ 4.5 \times 10^{-3} \mu m^{-1}$  & Walk-off bandwidth \\
\hline
    $q_{\rm D}$ & $\sqrt{ \frac{k_1}{l_c} }  $   & $ 5 \times 10^{-2} \mu m^{-1}$  
						&   Diffraction bandwidth@$\frac{\omega_0}{2}$   \\
\hline
    $q_{0, \rm D}$ & $\sqrt{ \frac{k_0}{l_c} }  $    & $ 7 \times 10^{-2} \mu m^{-1}$  
							& Diffraction bandwidth@$\omega_0$ \\
    \hline
    \end{tabular}
  \end{center}
 \caption{The table reports symbols and formulas for the relevant bandwidths introduced in the text, and their numerical values for conversion processes occurring between $527.5 \, {\rm nm}$  and $1055\, $nm in  a 4 mm BBO crystal. }
\label{table}
\end{table}
For upconversion of ultrashort pulses in a few mm crystal, this condition has been shown \cite{oshea2001}  to be satisfied easily,  unless the input pulse is as short as an optical cycle.  In our spatio-temporal description we need to add a similar condition on the effect of spatial diffraction: 
\beqa
\frac{q_{max} }{q_{\rm D}}   &\ll& \pi \quad  q_{\rm D} := \sqrt{ \frac {k_1}  {l'_c} } \label{qmax}
\eeqa
where we introduced a spatial bandwidth $q_{\rm D}$ related to diffraction. Its inverse,  $l_{\rm D} \propto \sqrt{\lambda l_c'}$,  is a diffraction length, describing the spatial broadening of the input light due to  diffraction. Condition \eqref{qmax} can be read as a requirement that the minimum spatial variation of the input light $\approx \, 1/q_{max}$ is larger than the diffraction length, in order that the beam is not spatially blurred by diffraction during propagation in the SFG crystal\footnote {To be precise, conditions  similar to \eqref{omegamax}, \eqref{qmax} should be introduced also for  the bandwidth of the upconverted light} . 
\par
Let us come back to the specific case when the input light is the parametric fluorescence:  in this case conditions \eqref{omegamax} and \eqref{qmax} are not generally satisfied. As shown by the example of Fig.\ref{fig_inc}a,  unless  the nonlinear crystal is extremely short,  the whole emission bandwidth of PDC is much broader than the bandwidths associated with GVD and diffraction (see table \ref{table} for numerical values).  From a slightly different perspective, the correlation  and coherence times of the PDC light have been shown to be close to the optical cycle \cite{jedr2006, gatti2009, jedr2012b}. \\
Thus, we have to resort to some more subtle approximation for phase matching: indeed, by rewrinting the SFG phase mismatch in terms of the phase mismatch of the original PDC light, we will derive a result similar to Eq.\eqref{delta_linear},  but  valid over a much broader range of spatial and temporal frequencies. 
With this goal in mind, we separate the longitudinal $k$-vector component of the PDC light into its even and odd parts with respect to the argument $\w= (\q,\Omega)$.
The even part can be expressed in terms of the  phase-matching function (\ref{delta_PW}) inside the PDC crystal: 
\beq
\frac{k_{1z}(\vka)+k_{1z}(-\vka)}{2}
=
\frac{\DeltaPDC(\vka)}{2}+\frac{ k_0^{\rm PDC}}{2}\, .
\label{k1z_even}
\eeq
For the sake of clarity, from now on we shall indicate by $k_0^{\rm PDC}$ and $k_0^{\rm SFG}$ the wave-number of the wave at frequency $\omega_0$ in either the PDC or SFG crystal, which can differ because the two crystal can be tuned at different angles. 
The odd part of $k_{1z}$ can be expanded in odd powers of $q$ and $\Omega$
\beq
\frac{k_{1z}(\vka)-k_{1z}(-\vka)}{2}
=
k_1'\Omega+\frac{1}{6}k_1'''\Omega^3+\frac{k_1'}{2k_1^2}q^2\Omega\label{k1z_odd}
+\hdots
\eeq
By neglecting the {\em cubic and higher order}  terms in Eq.(\ref{k1z_odd}) (starting from the third order dispersion and the term $\propto q^2\Omega$), the SFG phase-matching function can be recast as: 
\beq
\DeltaSFG (\vka,\vka_0-\vka)
\approx
{\cal D} (\vka_0)
+
\frac{1}{2}
\left\{
\DeltaPDC(\vka)+\DeltaPDC(\vka_0-\vka)
\right\},
\label{pm_incoh}
\eeq
with
\beq
{\cal D}(\vka_0)=k_0^{\rm PDC}-k_{0z}(\vka_0)+ k_1'\Omega_0\,.
\label{Dinc}
\eeq
This result is very useful because it allows us to rewrite the incoherent SFG spectrum (\ref{S_PWPA})
in the form
\begin{widetext}
\beqa
{\cal I}_{\rm SFG}(\vka_0)
&\propto&
 \int\frac{d\vka}{(2\pi)^{3}}
 {\cal I}_{\rm PDC}(\vka_0-\vka){\cal I}_{\rm PDC}(\vka) \,  
\sinc^2 \left\{ \frac{l_c'}{2}  {\cal D} (\vka_0 ) + \frac{l_c'}{4}  \left[  
\DeltaPDC(\vka) +\DeltaPDC(\vka_0-\vka)
\right]
\right\}  
 \label{Sincoh2}\\
&\approx& 
\sinc^2\left[\frac{{\cal D}(\vka_0 )l_c'}{2}\right]
\int\frac{d\vka}{(2\pi)^{3}}
 {\cal I}_{\rm PDC}(\vka_0-\vka){\cal I}_{\rm PDC}(\vka)
\label{Iapprox}
\eeqa
\end{widetext}
The last expression involves a somehow rough approximation,  based on the observation that  
the presence of ${\cal I}_{\rm PDC}(\vka)$ and ${\cal I}_{\rm PDC}(\vka_0-\vka)$ under the integral in Eq. (\ref{Sincoh2}) forces
$\DeltaPDC(\vka) \approx 0$ and $\DeltaPDC(\vka_0-\vka)\approx 0$ in the whole integration region,
due to the fact that the PDC spectrum is strongly peaked around the region where phase-matching occurs.
Strictly speaking it should be valid only in the limit $l_c\gg l_c'$, however it turns out to give a good description
of the SFG spectrum also when the two crystals have similar lengths.
\par
Within the validity of this approximation, the SFG spectrum takes thus the factorized form
\beqa
\label{ISFG}
&&
{\cal I}_{\rm SFG}(\vka_0)
\approx \cW(\vka_0)\cV(\vka_0).
\eeqa
where the first factor
\beqa
&&\cW(\vka_0)=(\sigma l_c')^2\sinc^2 \frac{{\cal D}(\vka_0)l_c'}{2},\label{cW}
\eeqa
describes the filtering effect due to phase matching in the SFG crystal and for a thick SFG crystal determines the overall shape of spectrum.
The second factor, which  does not depend on the SFG crystal parameters, 
\beqa
&&\cV(\vka)=
\int\frac{d\vka}{(2\pi)^{3}}
 {\cal I}_{\rm PDC}(\vka-\vkap){\cal I}_{\rm PDC}(\vkap)\;\;\;\;\;
 \label{cV}\,
\eeqa
is the self-convolution of the spectrum of the input PDC light, given by Eq.(\ref{PDC_spectrum}).   
\par
Therefore,  within this approximation the phase matching in the SFG crystal is effectively substituted by a function that depends only on the transverse  wave-vector and frequency of the upconverted light:
\beqa
 \DeltaSFG(\w, \w_0 - \w) & & \longrightarrow {\cal D} (\w_0) \\ 
 {\cal D} (\w_0) l_c' &=&
\left(k_0^{\rm PDC}-k_0^{\rm SFG} \right) l_c'  - \frac{\Omega_0} {\Omega_{\rm GVM}} + \frac {q_{0x} }{q_{\rm WO} }  \nn \\
&+& \frac{l_c' }{2k_0^{\rm SFG}} q_0^2  -\frac{k_0'' l_c' }{2} \Omega_0^2 +...
\label{delta_cubic}
\eeqa
where the definition \eqref{Dinc} and the espansion \eqref{k0z_quad} have been used to derive the second line. Formulas \eqref{ISFG}-\eqref{delta_cubic}
 represent the first main result of this work. Notice that up to linear terms Eq.\eqref{delta_cubic} looks similar to what obtained with the simpler linear approximation of phase matching \eqref{delta_linear},  with two remarkable differences: 
\\
i) Approximation \eqref{delta_cubic} is valid up to {\em quadratic terms},(it was obtained by neglecting third and higher order terms) and therefore is suitable to describe the upconversion of broadband PDC light.\\
ii) The zero order term in this formula is the difference between the wave numbers of the second harmonic in the two crystals, and is exactly zero when the two crystal are tuned for identical phase matching conditions, i.e. it could be zero even when both crystal are tuned for noncollinear phase matching. In comparison,  in formula \eqref{delta_linear}, the zeroth order term is the collinear phase mismatch in the SFG crystal. 
\par
As for the linear approximation \eqref{delta_linear},  the main contribution to phase matching comes here for the linear terms, which describe how the group velocity mismatch between the ordinary and extraordinary waves can be compensated by the spatial walk-off of the Poynting vector, resulting in a skewed geometry of the SFG spectrum that will be described in the next section. 
\subsection{The spectrum of upconverted PDC light} 
This section illustrates some example of the spectrum of the upconverted PDC light, calculated either with the plane-wave pump model or the stochastic simulations, and tests the validity of the approximations used to derive Eqs. (\ref{ISFG}-\ref{delta_cubic}). 
\par
\begin{figure*}[h,t]
\centering
{\scalebox{.7}{\includegraphics*{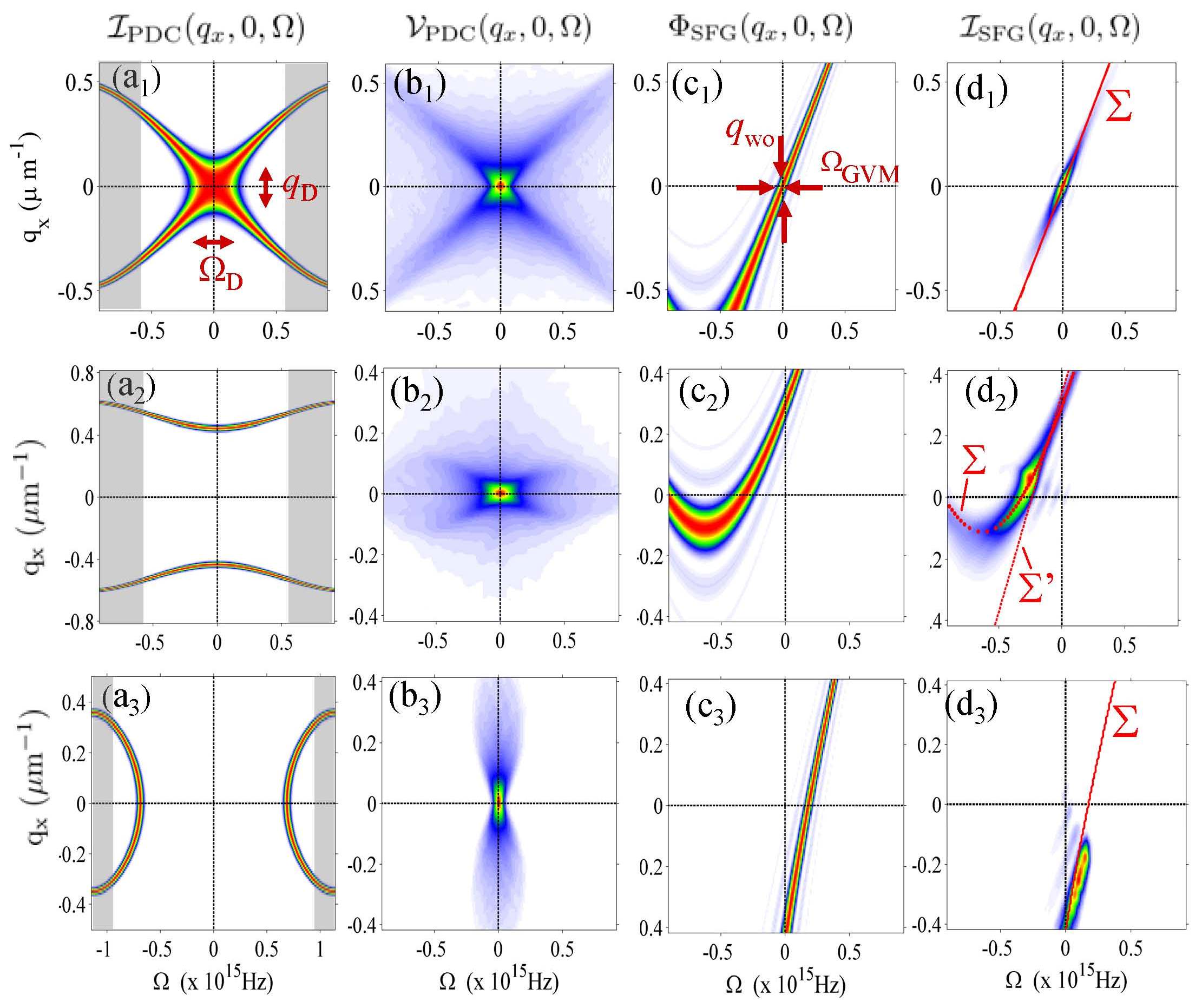}}}
\caption{(color online) Plots of (a) the PDC spectrum ${\cal I}_{\rm PDC}(\vka)$, (b) its self-convolution $\cV(\vka)$, (c) the effective phase-matching function in the SFG crystal $\cW(\vka)$, and (d) the upconverted spectrum ${\cal I}_{\rm SFG}(\vka)$ in the $(q_x,\Omega)$-plane, calculated from the plane-wave pump model. In all the plots the SFG crystal is tuned for collinear upconversion at $\theta_0^{\rm SFG}=22.9^{\circ}$, while in the first row the PDC crystal is tuned for collinear emission at $\theta_0^{\rm PDC}=22.9^{\circ}$, in the second row for non-collinear emission  $\theta_0^{\rm PDC}=23.9^{\circ}$, 
in the third row for nondegenerate emission $\theta_0^{\rm PDC}=21.9^{\circ}$.  The gray zones indicate the frequency filter used in the calculation to select a temporal PDC bandwidth of $\sim 10^{15}$Hz.
Other parameters: $l_c=4000\,\mu$m, $l_c'=1000\,\mu$m, $g=9.3$.}
\label{fig_inc}
\end{figure*}
Fig.\ref{fig_inc} describes the effect of various tuning conditions of the first PDC crystal (namely collinear,noncollinear and non-degenerate in the three rows), while the second crystal is kept fixed.   The first column  displays the spatio-temporal emission spectra of the input PDC light (the diffraction and GVD bandwidth are  indicated in the first plot). The second column shows the self-convolutions of such  spectra, that is,  the function $\cV(\vka)$ appearing in the SFG spectrum in Eq.(\ref{ISFG}). Column (c)  displays the filtering effect of phase matching in the second crystal due to the function $\cW(\vka)$. 
Finally,  the last column  plots the result for the  spectrum of upconverted light, evaluated from the full expression (\ref{S_PWPA}), without using the factorized approximation (\ref{ISFG}). 
It indeed demonstrates that the SFG spectrum is roughly the product of the functions $\cV$ and $\cW$, in agreement with
Eqs.(\ref{ISFG})-(\ref{cV}) (this result has been quantitatively verified), and that the shape of the spectrum is mainly determined by the function $\cW(\vka)$, i.e. by the phase matching in the second crystal.  
\par
Regarding this last point, we notice that within the approximations (\ref{Iapprox}-\ref{delta_cubic}) phase matching in the upconversion of PDC light is effectively achieved close to the surface $\Sigma$ in the 3D Fourier space defined by:   
\beq
\label{Sigma}
\Sigma:
{\cal D}(\q,\Omega)=0.
\eeq
where ${\cal D}(\q,\Omega)$ is the effective phase matching function given by Eq.\eqref{delta_cubic}. Notice that unless the crystal is very short, 
the GVM and walk-off bandwidths are much smaller than the dispersion and diffraction bandwidths, 
$\Omega_{\rm GVM} \ll \Omega_{0,\rm GVD}= 1/\sqrt{k_0'' l_c'} $ and $q_{\rm WO} \ll q_{0, \rm D}= \sqrt{k_0/ l_c'}   $ 
(See table \ref{table} for numerical values in a 4 mm crystal).  
Thus in a large central part of the spectrum  the linear terms in Eq.(\ref{delta_cubic}) describing temporal and spatial walk-off dominates over the quadratic terms, which can be neglected.  This part of the upconversion spectrum is thus close to the skewed plane determined by the equation 
\beq
\label{Sigma_prime}
\Sigma':
\frac{\Omega}{\Omega_{\rm GVM}} =
\frac{q_x}{q_{\rm WO}} + (k_0^{\rm PDC}-k_0^{\rm SFG})l_c'
\eeq
In these conditions,  relation (\ref{Sigma_prime}) shows esplicitly that phase matching occurs for those spatio-temporal modes 
for which the GVM between the fundamental and the up-converted light is compensated by their spatial walk-off.
Notice that the constant term is non vanishing  when the two crystals are tuned
for different phase-matching conditions. Accordingly, due to the different tuning of the first crystal, in the panels (c1), (c2), (c3) of  Fig \ref{fig_inc} phase matching in the second crystal occurs along different surfaces $\Sigma$, which,  as a first approximation,  can be considered as shifted in the plane $(q_x,\Omega)$ one with respect to the other. 
\par
\begin{figure}
\centering
{\scalebox{.6}{\includegraphics*{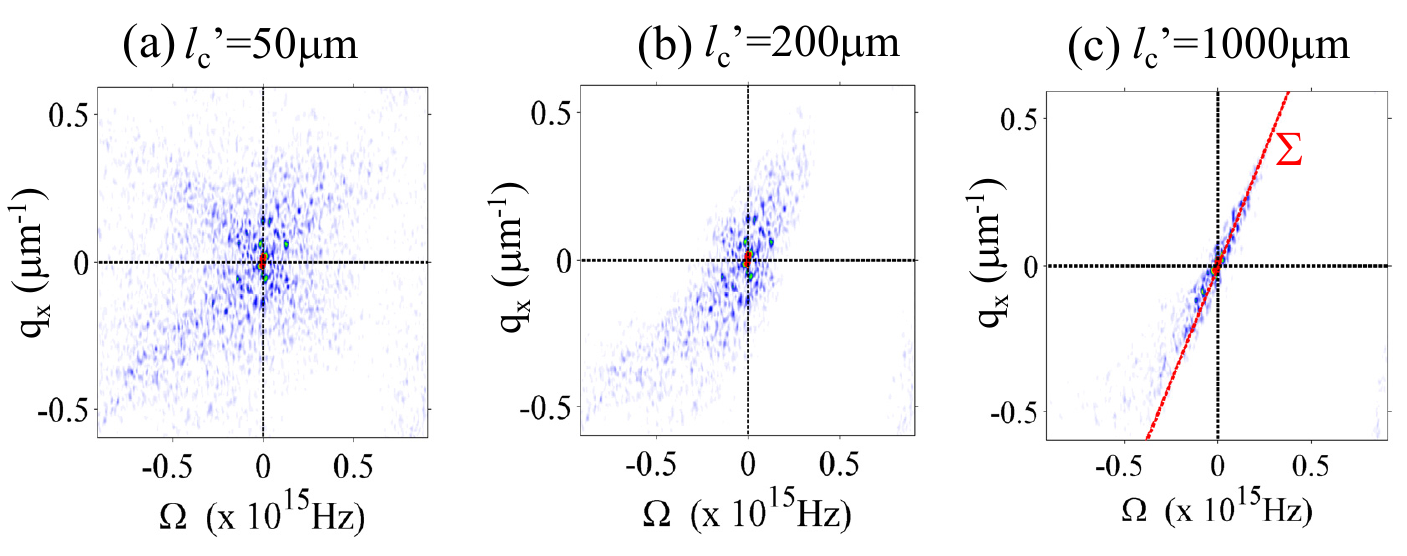}}}
\caption{(color online) Upconversion spectrum in the $(q_x,\Omega)$-plane for increasing lengths of the SFG crystal,  obtained from stochastic simulations of the propagation equations. Both crystals are tuned for collinear phase matching $\theta_0^{\rm PDC} = \theta_0^{\rm SFG} = 22.9^{\circ}$.  
The central peak at $(q_x=0,\Omega=0)$ is the coherent component. 
For $l_c'\ge 300 \mu$m the speckled incoherent spectrum from a single pump shot displays a distribution similar to the plane-wave-pump predictions ( Fig.\ref{fig_inc}d$_1$). $\tau_p=1$ps, $w_p=500{\rm \mu}$m, $g=9.3$,
$l_c=4000{\rm \mu m}$.}
\label{fig_spectr3D}
\end{figure}

We remark that this phase matching mechanism becomes effective in determining the geometry of the upconversion spectrum only when the length of the SFG crystal exceeds some hundreds of microns. This is illustrated by Fig. \ref{fig_spectr3D}, which shows how the SFG spectrum changes for increasing lengths of the SFG crystal $l_c'$, These plots are obtained by stochastic numerical simulations of the  propagation equations (\ref{prop1}-\ref{prop2}) in the two crystals, similar to those described in \cite{brambilla2012, jedr2012a, jedr2012b}. Parameters of the simulations are close to the experimental conditions, i.e. the pump driving the PDC has a waist $w_p=500\mu$m  and duration $\tau_p=1$ps, and the PDC gain $g\approx 9$ corresponds to a pulse energy close to $350\,{\rm \mu J}$. We verified that in these conditions the results of the numerical model are very similar to those of the plane-wave pump model. 
Fig. \ref{fig_spectr3D} plots the results of single stochastic realizations of the propagation equations, so that   the upconverted spectrum 
appears as a speckle-like pattern. \footnote{Ensemble averages performed by repeating the simulation would provide the correct quantum mechanical mean values in the the Wigner representation as
described in \cite{brambilla2004}. However, the required CPU time would be prohibitive and would not add more insight to the description.}
\par
For a thin SFG crystal ( Fig.\ref{fig_spectr3D}a), 
both spatial walk-off and GVM in the SFG crystal are almost ineffective, so that 
the SFG spectrum reproduces the  self-convolution of the PDC input spectrum (\ref{cV}), as can  be inferred by comparing Fig.\ref{fig_spectr3D} and fig.\ref{fig_inc}b$_1$.
By increasing $l_c'$, we see that the shape of the spectrum become close to the skewed plane defined by Eq.(\ref{Sigma_prime}), which corresponds to the modes for which  compensation of GVM by walk-off occurs.  
The simulations include also the coherent component of the SFG spectrum, visible as a  narrow peak centered at $(\q=0,\Omega=0)$ (notice that the plots are not in scale, the coherent peak is actually much larger than the incoherent component). 
\par
\begin{figure}   [h]
\centering
{\scalebox{.5}{\includegraphics*{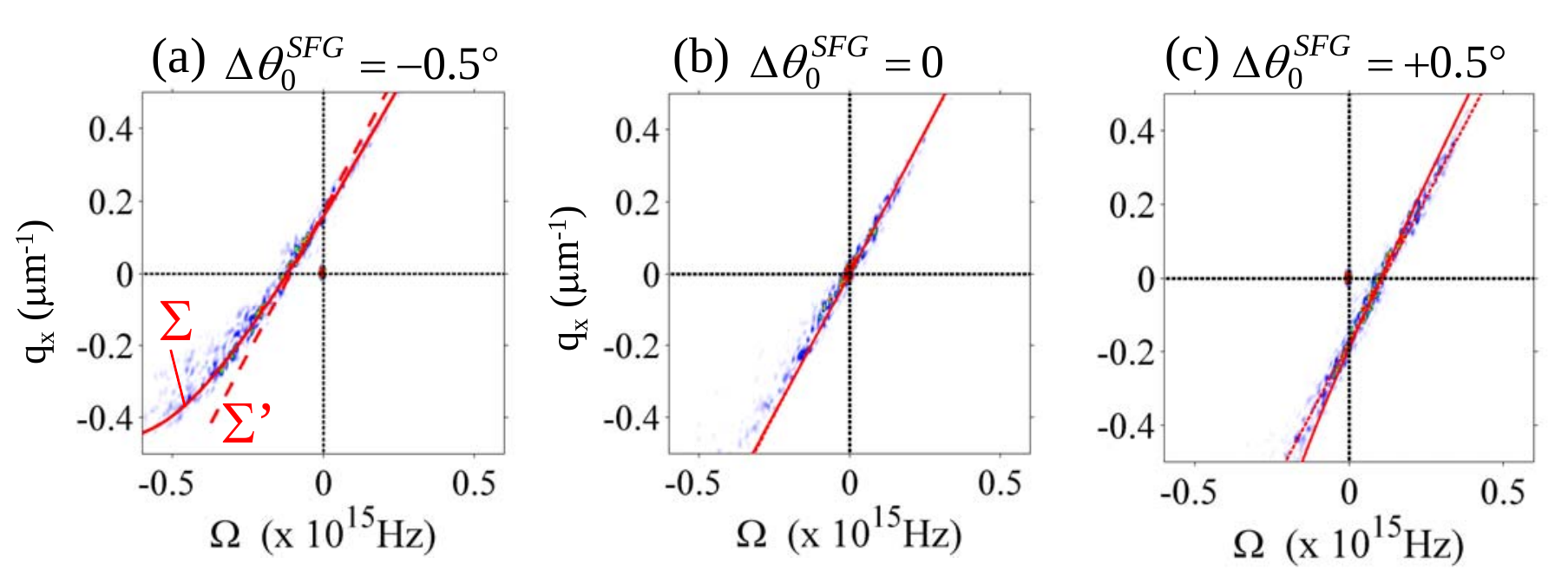}}}
\caption{(color online)SFG spectrum in the $(q_x,\Omega)$-plane (parallel to the walk-off plane) for different values of $\Delta\theta_0^{\rm SFG}$, 
obtained from a single realization of the stochastic simulations.  The solid and dashed red lines corresponds to the phase matching surfaces $\Sigma$  and $\Sigma'$, evaluated from the plane-wave pump model  Eq.(\ref{Sigma}) and Eq.(\ref{Sigma_prime}),  respectively. $l_c'=l_c=4000\mu$m, $g=9.3$, $w_p=500\,{\rm \mu}$m, $\tau_p=1$ps. }
\label{fig_spectrX}
\end{figure}
\begin{figure}[h]
\centering
{\scalebox{.5}{\includegraphics*{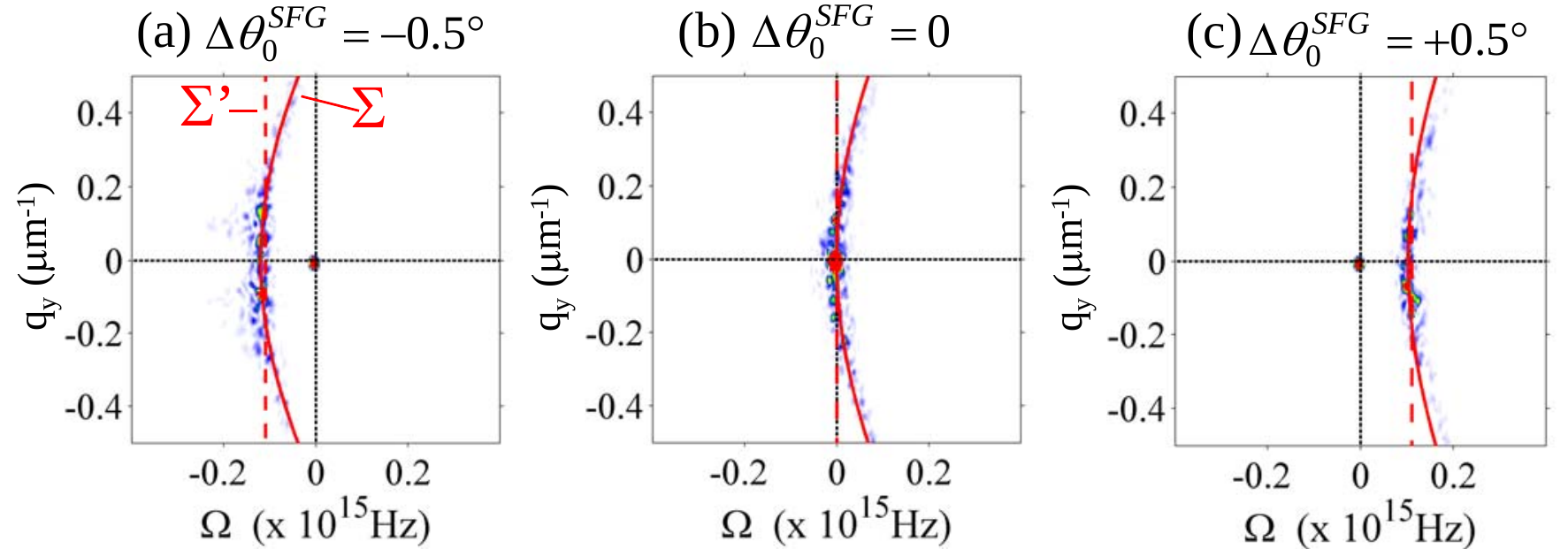}}}
\caption{(color online) SFG spectrum in the $(q_y,\Omega)$-plane (orthogonal to the walk-off plane) for the same parameters in Fig.\ref{fig_spectrX}}
\label{fig_spectrY}
\end{figure}

The effect of a different tuning of the two crystals on the phase matching mechanism can be also appreciated when the first crystal is kept fixed and the second one is tilted, as will be done in the experiment. This is shown in Fig. \ref{fig_spectrX} which plots the SFG spectrum in the plane parallel to the walk-off direction for different tilting angles of the SFG crystal, when the first crystal is tuned for collinear emission. In these plots the spectra appear skewed in the $(q_x,\Omega)$-plane along the plane $\Sigma'$ (red dashed lines in the plots), as predicted by the plane-wave pump model model [ Eq. (\ref{Sigma_prime})]. The red solid lines in the plots show the more precise phase matching surface given by Eq.(\ref{Sigma}),
which includes the slowly varying quadratic terms. The latter are responsible of
a slight concavity of the surface $\Sigma$ toward shorter wavelengths. 
Fig. \ref{fig_spectrY} shows the projection of the same spectra in the plane orthogonal to the walk-off direction, where the surface $\Sigma'$ appears as a vertical plane in the $(q_y,\Omega)$ plane.   As will be illustrated in the next section, Figs.  \ref{fig_spectrX} and \ref{fig_spectrY} corresponds to the observations  
in the the two possible configurations of the imaging spectrometer  implemented in the experiment. 
In the case of Fig.\ref{fig_spectrY} , since there is no spatial walk-off in the $y$ direction, upconversion occurs for all the transverse wave-vectors $q_y$ roughly at the same frequency,  whose offset from $\omega_0$ is given by
\beq
\label{omegaC}
\Omega_{\rm C} \approx  \Omega_{\rm GVM} \rho_0 l_c'  k_0^{\rm PDC}  \Delta\theta_0^{\rm SFG}
\eeq
as can be inferred by setting $q_x=0$ into Eq.(\ref{Sigma_prime}) and writing $ k_0^{\rm SFG} \approx k_0^{\rm PDC} - \rho_0  \Delta\theta_0^{\rm SFG} $, where the walk-off angle $\rho_0$ is defined by Eq. \eqref{rho0}. 
We notice a  symmetry with the situation where the two crystals are identically tuned $k_0^{SFG}= k_0^{PDC}$, and one looks at the SFG emission at different angles $\alpha_x$ in the walk-off plane. Setting  in this latter case $q_x= k_0^{SFG} sin \alpha_x \approx k_0^{PDC} \alpha_x$  in Eq. \eqref{Sigma_prime}, one obtains  : 
\beq
\Omega \approx \Omega_{\rm GVM} \frac{k_0^{PDC} \alpha_x} {q_{\rm WO} } = \Omega_{\rm GVM} k_0^{PDC} \rho_0 l_c'  \alpha_x 
\eeq
which coincides with the former equation \eqref{omegaC}, provided that the the propagation angle $\alpha_x$ is substituted by 
the tilt angle of the  crystal itself. This symmetry, derived here for small angles,  is indeed quite natural.
\par
We conclude this section by remarking  that the predicted behaviour of the incoherent upconversion spectrum versus a mistuning between the two crystals could be useful in order to optimize interferometric  measurements of twin beams realized by means of the SFG process,as those described in Refs.\cite{jedr2012a, jedr2012b},     
where the coherently upconverted component should be probed with high sensibility.  
As shown e.g by Figs.\ref{fig_spectrX}a,c a small tilt angle between the two crystals is enough to displace the whole spectrum of incoherent emission away from the coherent component, enhancing thus the visibility of its measurement.
\section{Experimental results}
\label{sec:exp}
The aim of the experimental work was to characterize the spatio-temporal far-field spectrum (in the $(\lambda,\alpha)$ plane) of the SFG radiation, and to verify the 
predicted behaviour of the incoherent component as a function of the angular mistuning between the PDC  and the SFG crystals. To this end the output radiation from the SFG crystal was analyzed by means of an imaging spectrometer (grating with $600\,$lines/mm), whose entrance vertical slit (with respect to the optical bench) was placed in the Fourier plane of a $20\,$cm focal length lens. The spectra were recorded in the walk-off plane and in
the plane orthogonal to walk-off.
Because of the geometry of our system and the crystal axis orientation, the vertical plane in the laboratory frame corresponded to the plane orthogonal to walk-off. 
In order to detect the spectra in the walk-off plane, the radiation reconverted from the SFG crystal was tilted by 90 degrees by means of two mirrors aligned in a periscope configuration.
\begin{figure}
\centering
{\scalebox{.5}{\includegraphics*{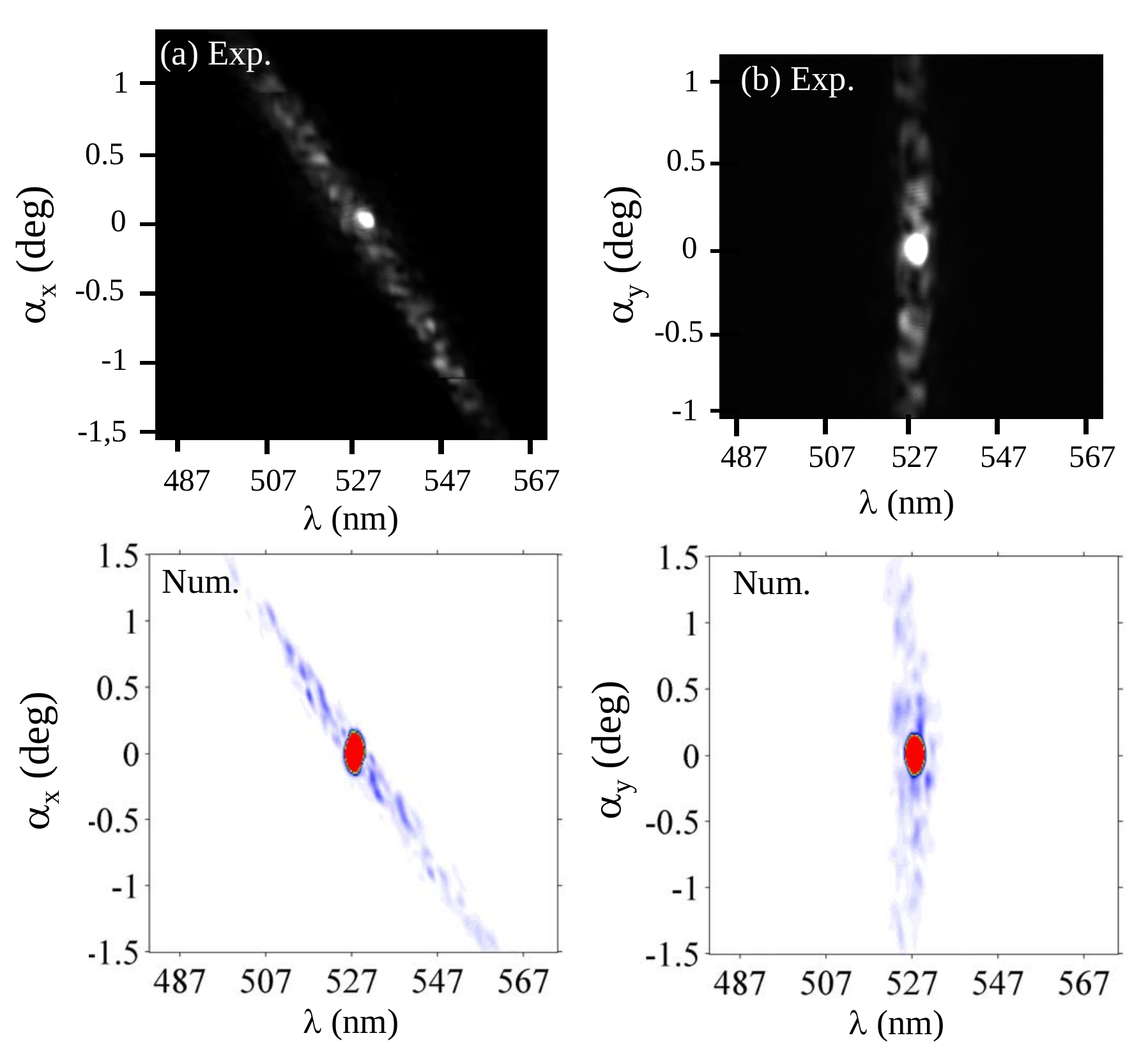}}}
\caption{(color online) Top: SFG far-field spectrum measured from a single pump shot with the IS slit selecting photons emitted (a) in the walk-off direction
and (b) orthogonal to the walk-off direction. The two crystals are perfectly tuned with $\Delta\theta_0^{\rm SFG}=0$.
Bottom: results of the numerical simulation implemented with the same parameters of the experiment.}
\label{fig_1shot}
\end{figure}
The upper panel reported in Fig.\ref{fig_1shot} (gray scale plots) shows typical single shot spectra of the SFG radiation recorded (a) in the plane parallel to walk-off and (b) in the orthogonal plane. The central coherent peak in the collinear direction is evident, and the features of the incoherent component in the $(\lambda,\alpha)$-plane are in agreement with the theory. The bottom panel shows the corresponding spectra obtained with the 3D numerical model. 
The experimental spectra exhibits the skewed geometry predicted by theory, which originates from the compensation of spatial
walk-off and GVM.
\par
\begin{figure}
\centering
{\scalebox{.5}{\includegraphics*{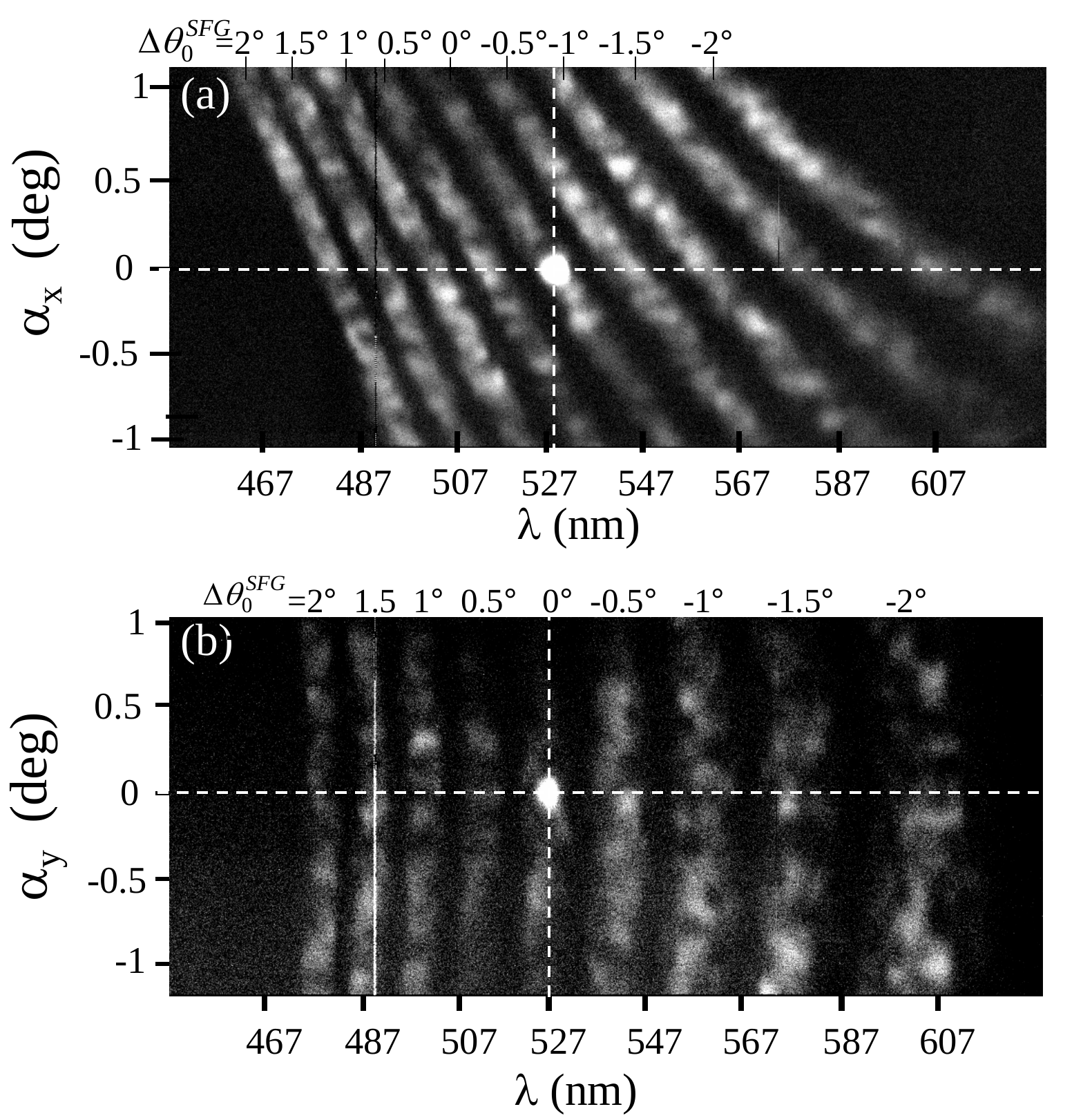}}}
\caption{SFG far-field spectrum measured with the imaging spectrometer for different tuning angles $\Delta\theta_0^{\rm SFG}$ of the SFG crystal (the values are indicated at top of the figure). The SFG radiation is recorded by the CCD after integration over 5 laser pump shots for each value of $\Delta\theta_0^{\rm SFG}$.}
\label{fig_XYplane}
\end{figure}

Figure \ref{fig_XYplane} illustrates the behaviour of the SFG spectrum when the second crystal is rotated with respect to the first one by an
angle $\Delta\theta_0^{\rm SFG}$; in particular it reports the SFG spectrum in the two different configurations by varying the tuning angle
from $-2^{\circ}$ up to $+2^{\circ}$ by steps of $0.5^{\circ}$. 
\par
\begin{figure}
\centering
{\scalebox{.7}{\includegraphics*{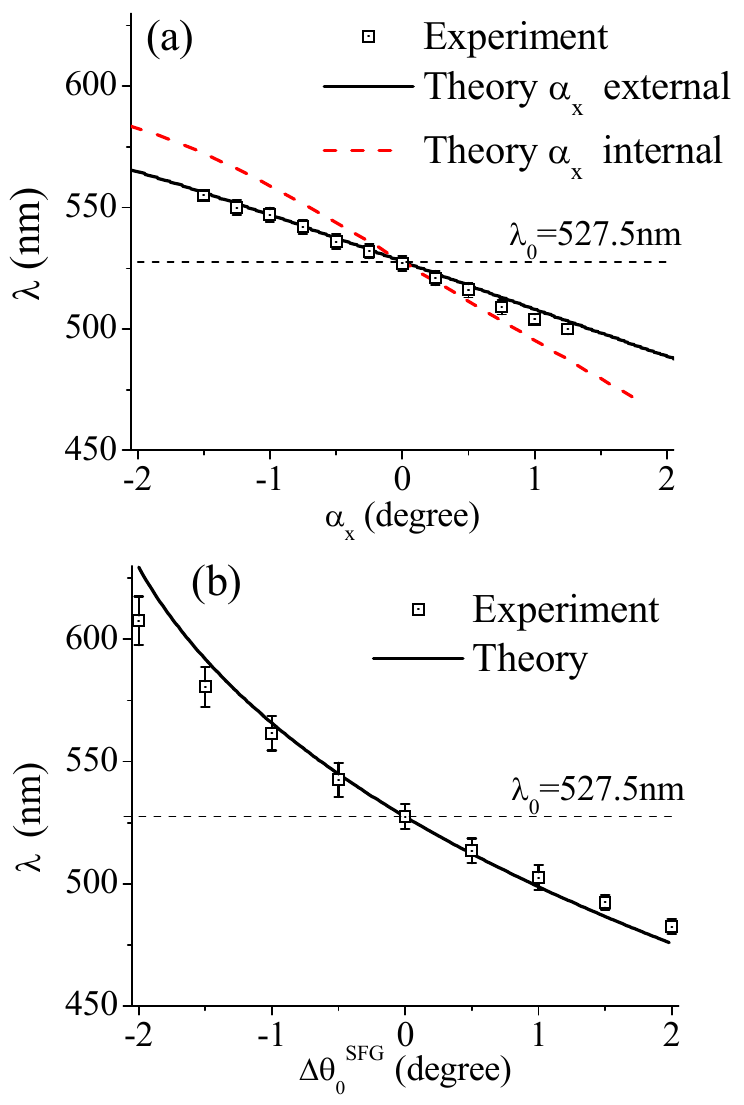}}}
\caption{Comparison between theory and experiment. (a)Up-converted wavelength versus the transverse angle of propagation $\alpha_x$, for a fixed angular tuning of both crystals at $22.9^\circ$.   Solid line and squares: theory and experimental results as a function of the external angle. Dashed line: theory as a function of  the internal angle. (b)Up-converted wavelength 
along the collinear direction as a function of the angular mistuning $\Delta\theta_0^{\rm SFG}$ between the two crystals.}
\label{fig_fit}
\end{figure}

Fig.\ref{fig_fit} displays a more detailed comparison between the theory and the experiment. Panel a)  reports 
the wavelength of the upconverted light (more precisely the peak value of the measured incoherent spectrum) as a function of the transverse propagation angle of the generated light, by keeping fixed the angular tuning of the two crystals.  Panel b) instead concentrate on a fixed propagation direction, namely the collinear z-direction $\q=0$ , and  the wavelength of the upconverted light is plotted as a function  of the angular mistuning between the two crystals. 
The theoretical curves are obtained form the plane-wave pump model,  by solving numerically the equation Eq.(\ref{Sigma}), 
${\cal D}(\q,\Omega)=0$, with ${\cal D}^{\rm inc}(\vka)$ being given by Eq.(\ref{Dinc})] . Precisely 
in the case of Fig.\ref{fig_fit}a $\Delta\theta_0^{\rm SFG}=0 $ and the equation is solved with respect to $q_x$ and $\Omega$ for $q_y=0$, while in Fig.\ref{fig_fit}b  $\q=0$ and the equation is solved with respect to $\Omega$ for different values of  $\Delta\theta_0^{\rm SFG} $.
The agreement between theory and experiment is very good although not perfect in plot b, especially at large mistuning angles. However,
such a quantitative agreement between theory and experiment is beyond the scope of this work, whose goal is mainly to
demonstrate the characteristic geometry of the spectrum of the SFG light originating from incoherent upconversion of PDC light.
\section{Conclusions}
In this work we investigated the  up-conversion of broadband PDC radiation, focusing on the spatio-temporal spectral properties  of the sum-frequency generated light.
\par
In particular, we have shown that propagation along the nonlinear crystal has the effect of selecting the upconverted spatio-temporal modes in a non-trivial way. The selection mechanism originates from the phase matching, which requires that the group velocity mismatch term is compensated by the spatial walk-off term. As a result, the different upconverted  frequencies propagate at different angles, and the upconverted spectrum is characterized, at least in  its central part, by  a skewed geometry in the space-time spectral domain.\\
 We provided a short comparison with the upconversion of an ultrafast coherent pulse in a thick crystal, where a similar mechanism is known to determine an angular dispersion of  the temporal frequencies. 
However, we have shown that the linear approximation of phase matching, valid in the case of an ultrafast input pulse, is not adequate in order to describe the spectrum of upconversion of the extremely broadband light generated by PDC:  
a complete characterization of this spectrum requires that  terms up to the quadratic ones are taken into account, and  shows that the shape of the upconversion spectrum depends drastically on the angular mistuning between the two crystals. 
We studied in detail the behaviour of the SFG spectrum with respect to such an angular mistuning, and   
demonstrated a progressive displacement of the incoherent spectrum with respect to the wavelength of the coherent peak for increasing values of the angular mistuning,  a property which could provide a useful tool for optimizing the visibility of twin beam correlation measurements. 
\par
These results have been illustrated by means of a semi-analytical and a numerical model of the optical system, 
and have been confirmed by the experimental measurements.
\par
The features above described, 
as for many other nonlinear optical processes, are a manifestation of the space-time coupling that occurs because of phase-matching.
We can cite the related examples of the skewed coherence along space-time trajectories predicted in three and four-wave mixing pr\cite{picozzi2002},
the X-shaped spatio-temporal coherence \cite{jedr2006} and quantum correlation \cite{gatti2009,caspani2010,brambilla2010,jedr2012a,jedr2012b} of twin beams 
and the macroscopic X-waves generated in quadratic media 
\cite{ditrapani2003} or in four-wave mixing processes \cite{couairon2006}. 
At the origin of these examples is the compensation of temporal dispersion by spatial diffraction.
In the present work the relevant mechanism  is instead the compensation of group velocity mismatch by spatial walk-off.
In a completely different context, the same mechanism has been exploited for the generation
of temporal solitons with a tilted pulse \cite{ditrapani98}.
\acknowledgments
This work was realized in the framework of
the Fet Open project of EC 221906 HIDEAS.

\end{document}